\documentclass[aps,prd,amssymb,showpacs,floatfix,twocolumn,superscriptaddress,nofootinbib]{revtex4-1}
\usepackage{amsmath,amsfonts,graphics,epsfig}

\begin{document}
\title{Pion gravitational form factors in a relativistic theory of
composite particles.}

\author{A.F.~Krutov}
\email{krutov@ssau.ru}, \affiliation{Samara State Technical University, 443100 Samara,
Russia},\affiliation{P.N.~Lebedev Physical Institute of the Russian Academy of Sciences, 443011 Samara, Russia},
\author{V.E.~Troitsky} \email{troitsky@theory.sinp.msu.ru}
\affiliation{D.V.~Skobeltsyn
Institute of Nuclear Physics,\\
M. V. Lomonosov Moscow State University, Moscow 119991, Russia}
\date{\today}
\begin{abstract}

We extend our relativistic theory of electroweak properties of
composite systems to describe simultaneously the gravitational form factors
of hadrons. The approach is based on a version of the instant-form
relativistic quantum mechanics and makes use of the modified impulse
approximation. We exploit the general method of the relativistic
invariant parametrizaton of local operators to write the
energy-momentum tensor of a particle with an arbitrary spin.
We use
the obtained results to calculate the gravitational form factors of the
pion assuming point-like constituent quarks. All but
one parameters of our first-principle model were fixed previously in
 works on electromagnetic form factors. The only free parameter, $D_{q}$,
is a characteristic of the gravitational form factor of a constituent
quark. The derived form factors of the pion satisfy the constraints given
by the general principles of the quantum field theory of hadron structure.
The calculated gravitational form factors and gravitational mean-square
radius are in a reasonable agreement with known results.

\end{abstract}

\maketitle

\section{Introduction}
\label{sec: Intr}

The probably most   fundamental information about a particle is contained
in the matrix elements of its energy-momentum tensor (EMT).
So, it is clear that the
gravitational form factors (GFFs) of hadrons that enter the EMT matrix
elements and their dependence on the square of the momentum transfer $t$
are in the focus of investigations (see, e.g. \cite{LeL14, Ter16, PoS18,
CoL20, BuE18} and references therein). These form factors contain the
information about the distribution of mass, spin and internal forces
inside the hadron. These forces are connected with the additional global
characteristic of a particle, the so-called $D$ term of the EMT matrix.
The study \cite{LoC17} of the structure of the elements of the EMT matrix
and of their Lorentz-covariant decomposition in terms of the form factors
gives the limitations at $t \to 0$ that mean that the form factor $D(t)$
is not constrained (not even at $t = 0$) by general principles and the
value $D = D(0)$ is therefore not known for (nearly) any particle (see
also \cite{CoL19, LoL20}).

At present one obtains the information about the GFF
mainly from the  hard-exclusive processes described in terms of
unpolarized generalized parton
distribution (GPD). Particularly, in Ref. \cite{Pol03} (see also Ref.
\cite{PoS19}) it is shown that GPD, derived from processes,
gives the information on the space distribution of strong forces
that act on quarks and gluons inside hadrons.
The link of gravitational form factors with GPD gives a possibility of
obtaining the data on these form factors using the
hard-exclusive processes. The first results for nucleon GFFs were obtained
through the analysis of JLab data \cite{BuE18}. The data for the pion form
factors were extracted from the experiment of the collaboration Belle at
KEKB \cite{Mas16,KuS18}.

It is worth noting that the model-independent extraction of GPD from the
experimental data is a difficult long-term  problem. So, today the
theoretical estimation
of the GFF, including $D$-term, is usually
obtained in the framework of different model approaches.
We mention here only the publications that are strictly related to the
present paper while more general information can be found, for example,
  in the reviews
\cite{LeL14, Ter16, PoS18, CoL20}.

The pion electromagnetic and gravitational form factors are obtained from
GPD in the Nambu--Jona-Lasino (NJL) model
\cite{FrC19,FrC20}. The authors show, in particular,
that the light-cone mass  radii for the pion are almost twice smaller
than the light-cone charge radii (see also \cite{BrA08}).
The calculated light-cone mass radius
agrees with the value obtained through a phenomenological extraction from
KEKB data \cite{KuS18}. Note that the NJL model as well as the model of
our approach contains gluons only implicitly.

Different approaches based on various forms of dispersion
relations
(see, e.g., \cite{Pas18, PaP14} and references therein),
that is on quite general principles of the quantum-field theory,
can be considered as "maximal model-indepemdent"$\,$ approaches.
The $D$ term was calculated using unsubtracted $t$-channel
dispersion relations for the deeply virtual Compton scattering amplitudes
\cite{Pas18}.

In connection with the first experimental results for the GFFs of nucleons
\cite{BuE18}, the
gravitational characteristics of these particles were calculated. The
dependence of EMT on the long-ranged electromagnetic interaction was
investigated in Ref. \cite{VaS20}. Using a simple model it was shown
that in the case of the long-ranged forces in the proton one needs a
sophisticated theory of  the $D$-term construction. There is a possibility
that the $D$-term is ill-defined and even singular. They propose the
exploiting of the fixed-$t$ dispersion relations for deep
virtual  Compton scattering as in Ref. \cite{PoS19} to avoid this
difficulty. It is of interest that in the free-field model, as was shown
in Ref. \cite{HuS17, HuS18}, the $D$-term for the fermion with spin 1/2 is
of a dynamical nature and vanishes for the free fermion. The interaction
inclusion gives rise to the $D$-term of a fermion with an internal
 structure, the nucleon. Recently \cite{LoM19} the results of \cite{PoS18}
have been extended to the different frames where the nucleon has a
nonvanishing average momentum.

The authors of \cite{PeP16} use the Skyrme model
which respects the chiral symmetry and provides a practical realization of
the large-$N_{c}$ picture of baryons described as solitons of mesonic
fields. The EMT form factors are consistently described (see, e.g.,
\cite{NeS20} and references therein) in the bag model in the
large-$N_c$ limit. It is important to mention the recent results for the
GFFs as obtaibed from the lattice QCD (see, e.g., \cite{ShD19a,ShD19b} and
the references therein), and the study using an approach based on the
light-cone sum rules \cite{Ani19}, and in the light-cone quark model
\cite{SuD20}.

Different theoretical approaches to the GFFs of
hadrons give different results and the absence of the
model-independent extracted data makes it impossible to choose between
them. We believe that in such a situation, a
theory which is intrinsically self-consistent theory and  describes as
large set of the physical charathacteristics and systems as possible is
welcome. This motivates  the  extension
of the relativistic theory of the electromagnetic properties of composite
systems developed previously to the calculation of the GFF of the pion.

The goal of the present paper is twofold. First, we extend our
relativistic model of the electromagnetic structure of composite systems to
include their gravitational characteristics. Second, we derive the pion
GFFs using our
previous calculations of electroweak properties of hadrons.

The model
\cite{KrT02, KrT03} was successfully used for various composite
two-particle systems, namely, the deuteron \cite{KrT07}, the pion
\cite{KrT01, KrT98, KrT09prc, TrT13}, the $\rho$ meson \cite{KrP16,
KrP18, KrT19} and the kaon \cite{KrT17}. This model had predicted, with
surprising accuracy, the values of the form factor $F_{\pi}(Q^{2})$, which
were measured later in JLab experiments  (see the discussion in Ref.
\cite{KrT09prc}): all new measurements follpwed the predicted curve.
Another advantage of the approach is matching with the QCD predictions in
the ultraviolet limit, when constituent-quark masses are switched off, as
expected at high energies. The model reproduces correctly not only the
functional form of the QCD asymptotics, but also the numerical
coefficient; see Refs. \cite{KrT98, TrT13, KrT17} for details. The method
allows for an analytic continuation of the pion electromagnetic form
factor from the space-like region to the complex plane of momentum
transfers and gives  good results for the pion form factor in the
time-like region \cite{KrN13}.

Now we show that besides electroweak properties of composite systems, our
approach can be used to calculate their gravitational characteristics.
Even in a simple version of our approach (with the point-like
quarks and the two-particle wave functions of the harmonic oscillator)
the results agree well with other calculations and with scarce
measurements.  The only free parameter that we add to the model is the
constituent-quark $D(0)=D_{q}$. This parameter is constrained from the
pion mean-square radius. Despite uncertainties in the latter, $D_{q}$ is
fixed to a narrow interval which makes it possible to predict the GFFs at
nonzero momentum transfers. Using the obtained results we calculate the
values of the static gravitational characteristics of the pion and obtain
$A$ and $D$ form factors as functions of momentum transfer up to
1\,GeV$^2$. Note that the new parameter is not used in the calculation of
the $A$ term, its value is a direct prediction of our previous approach.
The form factors calculated through our nonperturbative method satisfy all
the constraints given by the general principles of the quantum-field
theory of hadron structure \cite{LoC17, CoL19, LoL20}.

The approach that we use is a particular variant of the theory
based on the classical paper
by P.~Dirac \cite{Dir49}, so-called Relativistic Hamiltonian
Dynamics or
Relativistic  Quantum Mechanics (RQM). It can be formulated in different
ways or in different forms of dynamics. The main forms are the instant
form, point form and light-front dynamics. Here we are dealing with
instant-form (IF) RQM. The properties of different forms of RQM dynamics
are discussed in the reviews \cite{LeS78, KeP91, Coe92, KrT09}. Today the
theory is largely used as a basis of the nonperturbative approaches to the
particles structure.

The presentation of the matrix elements of EMT in terms of form factors,
 the invariant parametrization, is an important part of model
approaches to gravitational characteristics of particles.
The majority of authors use the parametrization given by
Pagels \cite{Pag66} (see also \cite{KoO63}). The parametrization \cite{Pag66}
was constructed in an almost phenomenological way using an
analogy with the investigations performed in connection with the
self-stress of the electron. It is valid for the simplest cases of spin
$0$ and $1/2$ and can not be directly extended to systems with higher
spins, when more general special methods are needed (see
\cite{ChS63,CoL20}). We use the general method of the relativistic
invariant parametrization of the matrix elements of the local operators
established in Ref. \cite{ChS63}. The parametrization is written in
the canonical basis, so it is natural to call it the canonical
parametrization. Certainly, this method of obtaining a parametrization is
not unique (see the discussion in Ref. \cite{CoL20}). With the use of the
method \cite{ChS63} the invariant parametrization was obtained for systems
with arbitrary spin in the cases with diagonal \cite{KrT02, KrT03, KrT05}
and nondiagonal \cite{KrP15} total angular momenta. In the present paper we
give the general formulae, although for the actual calculation we use only
the form factors for systems of spin $0$(the pion), spin $1/2$ (the
costituent quark) and for the system of two free constituent quarks. The
form factors obtained by the canonical parametrization can be expressed in
terms of the largely accepted one-particle GFFs \cite{PoS18, Pag66}.

As an important part, our approach contains the construction of the EMT
matrix element of the system of two free particles with spins $1/2$,
momenta $\vec p_1\,,\,\vec p_2$ and spin projections
$m_1\,,\,m_2$, that is the two-particle system having quantum numbers of
the pion. We construct the  EMT in the basis with separated center-of-mass
motion \cite{KoT72} $|\vec P\,,\,\sqrt{s}\rangle\,$, where $P =
p_1+p_2\,,\,s = P^2\,$ is the invariant mass squared and $P$, $p_1$
and $p_2$ are 4-vectors. We refer to the corresponding form factors as to
the free two-particle GFFs. These form factors are the functions of the
invariant masses of the two-particle system in the initial and the final
states and depend on the momentun-transfer square as a parameter.
They are the regular generalized functions, the distributions
corresponding to the functionals given by the the two-dimensional
integrals over the invariant masses \cite{KrT02}.

To construct the pion GFFs we use
a modified impulse approximation (MIA) (see
Refs. \cite{KrT02,KrT03} and the review \cite{KrT09}).
 In contrast to the baseline impulse
approximation, MIA is formulated in terms of the form factors
and not in terms of the EMT operator itself.
So, in MIA the pion GFFs
are presented as functionals given by the free two-particle form factors
on the set of the two-quark wave functions of the pion.
\textit{The necessity} of using the distributions was justified
in the case of the electroweak interaction in
\cite{KrT02, KrT03, KrT05} (see also \cite{KrT09,LoC17}).

The rest of the paper is organized as follows.
In Sect.\ref{sec: Sec 2} we construct the matrix elements of the EMT of
the particle with an arbitrary spin and, in particular, with spins
$0$ and $1/2$.
Sect.\ref{sec: Sec 3} presents the construction of the EMT matrix element
for the system of two free spin $1/2$ particles with total quantum numbers
of the pion and the explicit forms of corresponding form factors.
In Sect.\ref{sec: Sec 4} we give a brief account of RQM and MIA and derive
the formulae for the pion GFFs. In Sect.\ref{sec: Sec 5} we discuss the
important role of the relativistic effects in the pion GFFs behavior. We
calculate the static limits of the GFFs and of their derivatives, obtain
the mean square radius and the values of the form factors up to
1\,GeV$^2$. We briefly conclude and discuss the results in Sect.\ref{sec:
Sec 6}.

\section{The energy-momentum tensor  matrix elements\\
for a particle with an arbitrary spin}
\label{sec: Sec 2}

In this Section we describe the general procedure of parametrization  of
the EMT matrix element for a particle with  mass
$M$ and spin $j$. To write EMT in terms of gravitational form factors  we
make use of the method
\cite{ChS63}. Because of translational invariance it is sufficient to
consider only the following matrix element:
\begin{equation}
\langle \vec p, m\left|T_{\mu\nu}(0)\right|\vec p\,',m'\rangle\;,
\label{Tj}
\end{equation}
where $\vec p\,',\,\vec p$ are the particle moments, $m'\,,\,m$ are the
spin projections in the initial and final states, respectively;
$p'\,^2 = p^2 = M^2$.

The normalization condition for the state vectors in (\ref{Tj}) is:
\begin{equation}
\langle \vec p, m\left|\right.\vec p\,',m'\rangle = 2\,p_0\,\delta(\vec p -\vec p\,')\delta_{mm'}\;,
\label{norm1}
\end{equation}
with $p_0=\sqrt{M^2 + \vec p\,^2}$.
We have exploited the general method of parametrization
of matrix elements of local operators developed in \cite{ChS63}
to construct the matrix elements of the operator of
the electromagnetic current (see, e.g., \cite{KrT02,KrT03,KrT09}). Upon
formulation of this method the canonical basis in the Hilbert space was
used.  From the point of view of group theory the parameterization
procedure represents the realization of the known Wigner -- Eckart
theorem on the Poincar\'e group \cite{KrT05}. The  parameterization
represents the procedure of separation of the reduced matrix elements
(form factors) which are invariant with respect to transformations of the
Poincar\'e group. The main idea of the canonical parameterization can be
formulated  as follows. Objects of two types should be constructed from
the variables in the vectors in the Hilbert space in (\ref{Tj}):

1. The set of linearly independent matrices in spin projections in  the
initial and final states. At the same time this set
represents the set of linearly independent Lorentz
scalars (scalars and pseudoscalars).  This set describes the EMT matrix
elements non-diagonal with respect to $m,m'$  and the behavior of the
matrix elements under discrete space-time transformations.

2. The set of linearly independent objects with the same tensor dimension
as the operator. In our case (\ref{Tj}) this is a 4-tensor of the rank two.
This set describes the behavior
 of the matrix elements under Lorentz transformations.

The matrix
element of the operator is written as the sum of all
possible products of objects of the first type and objects of the second
type. The coefficients of the elements of this sum are the desired reduced
matrix elements, that is form factors. The obtained linear combination is
modified if additional constraints, for example, conservation laws, are
imposed on the EMT operator.

To construct a Lorentz-invariant matrix in spin projections we use the
well-known 4-pseudovector of (see, e.g., \cite{Pol19}):
$$
\Gamma_0(p) = (\vec p\vec j)\;,\quad
\vec \Gamma(p) =  M\,\vec j + \frac {\vec p(\vec p\vec j)}{p_0 + M}\;,
$$
\begin{equation}
\Gamma^2 = -M^2\,j(j+1)\;.
\label{Gamma mu}
\end{equation}
Under the Lorentz transformations
$p^\mu = \Lambda^\mu_{\;\nu}\,p'\,^\nu$,
the operator of the 4-spin
 (\ref{Gamma mu})  is transformed according to the representation of the
small group:
\begin{equation}
\Gamma^\mu(p) = \Lambda ^\mu
_{\;\nu}\,D_w^j(p,\,p')\,\Gamma^\nu(p')\,D_w^j(p',\,p)\;,
\label{Lambda Gamma mu}
\end{equation}
where $\Lambda^\mu_{\;\nu}$ is the matrix of a Lorentz transformation and
$D_w^j(p,\,p')$ is the transformation operator from the small group, the
matrix of three-dimensional rotation. The Lorentz-transformation matrix in
our case is of the form
\begin{equation}
\Lambda ^\mu _{\;\nu} = \delta^\mu_\nu + \frac{2}{M^2}p^\mu\,p'_\nu -
\frac{(p^\mu + {p'}\,^\mu)(p_\nu + p'_\nu)}{M^2 + p^\lambda\,p'_\lambda}\;.
\label{Lambda}
\end{equation}
It can be shown using (\ref{Lambda Gamma mu}) that matrix elements of the
operator $D_w^j(p,\,p')\Gamma^\mu(p')$ transform as the
4-pseudovector and matrix elements of the operators
$D_w^j(p,\,p')p_\mu\Gamma^\mu(p')$ and $p'_\mu\Gamma^\mu(p)
D_w^j(p,\,p')$ as 4-pseudoscalars.
Thus, the set of linearly independent scalars composed of the vectors
$p^\mu,\>{p'}^\mu$ and the pseudovector $\Gamma^\mu(p')$ contains not
only diagonal (with
respect to spin projections) terms, but non-diagonal terms, too.
Note, that the pseudovector $\Gamma^\mu (p)D_w^j(p,\,p')$ does not enter
the set of scalars. Its linear dependence can be shown if
we use relation (\ref{Lambda Gamma mu}) and the explicit form of the matrix
$\Lambda ^\mu _{\;\nu}$ (\ref{Lambda}).
After simple calculations we obtain
$$
\Gamma^\mu (p)D_w^j(p,\,p') = D_w^j(p,\,p')\left[\Gamma^\mu (p') - \right.
$$
\begin{equation}
\left. - \frac {p^\mu
+{p'}^\mu}{M^2 + p_\mu {p'}^\mu}\,\left[p_\nu \Gamma^\nu (p')\right]\right]\;.
\label{Gamma D = D Gamma}
\end{equation}
Since ${p'}_\mu \Gamma^\mu (p') = 0$, the desired set of linear indepent
matrices (that is the set of independent Lorentz scalars) is given by
$2j + 1$ elements
\begin{equation}
D_w^j(p,\,p')\,(ip_\mu \Gamma^\mu (p'))^n\;,\quad n = 0,1,2\ldots ,2j\;.
\label{pseud}
\end{equation}
The imaginary unit $i^2=-1$ is introduced for self-adjointness of
the obtained scalar operators (\ref{pseud}).
 The self-adjointness property can be proved
using the relation following from (\ref{Gamma D = D Gamma}):
\begin{equation}
p'_\mu \Gamma^\mu (p)D_w^j(p,\,p') = -D_w^j(p,\,p')p_\mu \Gamma^\mu (p')\;.
\label{p Gamma D = - Dp Gamma}
\end{equation}
The number of linearly independent scalars in  (\ref{pseud}) is limited by
the fact that the product
containing more than $2j$ numbers of factors $\Gamma^\mu (p')$
is reduced to the products of smaller number of factors, i.e., is
not linearly independent.
For even $n$ the obtained objects in (\ref{pseud}) are
scalars, and for odd $n$ they are pseudoscalars.

In the decomposition of the matrix element
(\ref{Tj}) we make use of the metric pseudotensor
$g_{\mu\nu}$ and the rank 2 tensors, that should be constructed from
the variables on which the state vectors in (\ref{Tj}) do depend.
Using the available variables in the state vectors of the particle,
it is possible to construct one pseudovector
$\Gamma^\mu (p')$ (\ref{Gamma mu}) and three
independent vectors:
$$
K_\mu = (p - p')_\mu\,,\quad K'_\mu = (p + p')_\mu \,,
$$
\begin{equation}
R_\mu = \epsilon _{\mu \,\nu \,\lambda
\,\rho}\, p^\nu \,p'\,^\lambda \,\Gamma^\rho (p')\;.
\label{kk'RG}
\end{equation}
Here $\epsilon _{\mu \,\nu \,\lambda\,\rho}$
 is the absolutely antisymmetric pseudotensor of rank 4,
$\epsilon_{0\,1\,2\,3}= -1$.
For the matrix elements of the operators in
(\ref{kk'RG})
 to transform as the 4-vector,  it is necessary to multiply them by
$D_w^j(p,\,p')$
from the left (in analogy with (\ref{pseud})).

The matrix element (\ref{Tj}) is written in terms of all
possible products of vectors (\ref{kk'RG}), pseudovector $\Gamma_\mu (p')$,
and pseudotensor $g_{\mu\nu}$. Each of  these objects is multiplied by
a sum of linearly independent scalars (\ref{pseud}). The coefficients in
such a decomposition are just form factors, or reduced matrix elements.

Taking into account the symmetry properties of the EMT the parametrization
of the matrix element (\ref{Tj}) can be written in the form:
$$
\langle \vec p, m\left|T^{(\pi)}_{\mu\nu}(0)\right|\vec p\,',m'\rangle
= \sum_{m''}\,\langle\,m|D_w^j(p,\,p')|m''\rangle\times
$$
\begin{equation}
\times\langle\,m''|\tau_{\mu\nu}(0)|m'\rangle\;,
\label{T=tau}
\end{equation}
where
$$
\tau_{\mu\nu}(0) = G_1\,K'_\mu\,K'_\nu + G_2\,\Gamma_\mu\Gamma_\nu + G_3(K'_\mu\Gamma_\nu +
\Gamma_\mu\,K'_\nu) +
$$
$$
+ G_4\,(K'_\mu\,R_\nu + R_\mu\,K'_\nu) + G_5(R_\mu\Gamma_\nu + \Gamma_\mu\,R_\nu) +
$$
$$
+ G_6\,K_\mu\,K_\nu+ G_7\,g_{\mu\nu} + G_8(K_\mu\Gamma_\nu + \Gamma_\mu\,K_\nu) +
$$
\begin{equation}
+ G_9(K'_\mu\,K_\nu + K_\mu\,K'_\nu) + G_{10}(K_\mu\,R_\nu + R_\mu\,K_\nu)\;,
\label{tau}
\end{equation}
\begin{equation}
G_i = \sum_n\,g_{in}(t)(ip_\mu\Gamma^\mu(p'))^n\;.
\label{Gi}
\end{equation}
In (\ref{Gi}),  $g_{in}(t)$ are the invariant coefficients, form factors,
$t = K^2$ is momentum-transfer square and  $\Gamma_\mu =\Gamma_\mu(p')$.

Let us impose some additional physical conditions on the operator
(\ref{T=tau}).

1. The requirement of self-adjointness.
It is easy to show, making use of (\ref{p Gamma D = - Dp Gamma}), that
the self-adjointness for r.h.s. of (\ref{T=tau}) requires a modification
of the pseudovector $\Gamma_\mu$ with the help of the quantities
introduced by (\ref{pseud}), (\ref{kk'RG}); namely:
\begin{equation}
\Gamma^\mu\;\to\;\tilde\Gamma_\mu = \Gamma^\mu(p') - \left(\frac{K^\mu}{K^2}+ \frac{{K'}\,^\mu}{{K'}^2}\right)
\left[p_\mu\Gamma^\mu(p')\right]\;.
\label{mod}
\end{equation}
Note that this modification
(\ref{mod}) does not affect the Lorentz scalars (\ref{pseud}).
The requirement of self-adjointness also results in the
multiplication of the terms containing $G_4,G_5,G_8,G_9$ by the imaginary
unit.

2. The conservation law for EMT, $T_{\mu\nu} K^\mu = 0$, gives the
following conditions to be imposed on the Lorentz scalars,
\begin{equation}
G_8=G_9=G_{10} = 0\;.
\label{Gi=0}
\end{equation}
The conservation law requires also the following changes:
\begin{equation}
G_6\to\;-G_6\;,\quad G_7\to\;tG_6;.
\label{G67}
\end{equation}

3. The parity-conservation condition gives limitations for the summation in
 (\ref{pseud}). Namely, in  $G_1, G_2, G_4, G_6$ the values of $n$ are
 even while in $G_3, G_5$ they are odd. The limits of summations are the
 following: for $G_1,G_6$ they are $0\le n\le 2j$; for $G_3,G_4$ they are
 $0\le n\le 2j-1$; for $G_2,G_5$ they are  $0\le n\le 2j-2$.
Summing is limited by the fact that each term in the decomposition
 (\ref{tau}) contains no more than $2j$ factors $\Gamma(p')$.

So, the most general parameterization of the matrix element  (\ref{T=tau})
has the following form if the  above constraints are taken into account:
$$
\tau_{\mu\nu}(0) = \frac{1}{2}G_1\,K'_\mu\,K'_\nu + G_2\,\tilde\Gamma_\mu\tilde\Gamma_\nu + G_3(K'_\mu\tilde\Gamma_\nu +
\tilde\Gamma_\mu\,K'_\nu) +
$$
$$
+ iG_4\,(K'_\mu\,R_\nu + R_\mu\,K'_\nu) + iG_5(R_\mu\tilde\Gamma_\nu + \tilde\Gamma_\mu\,R_\nu) +
$$
\begin{equation}
+ G_6(tg_{\mu\nu} - \,K_\mu\,K_\nu)\;,
\label{taufin}
\end{equation}
where the summation is limited as is pointed above and the factor 1/2
 before $G_1$ is a result of the normalization condition: the static limit
 of EMT should be equal to the mass.

Let us use the obtained general parametrization in the case of spin 0.
Now for the pion EMT we have:
$$
\langle \vec p\left|T^{(\pi)}_{\mu\nu}(0)\right|\vec p\,'\rangle = \frac{1}{2}G^{(\pi)}_{10}(t)K'_\mu K'_\nu +
$$
\begin{equation}
+ G^{(\pi)}_{60}(t)\left[tg_{\mu\nu} - K_\mu K_\nu\right]\;,
\label{Tpi}
\end{equation}
The pion GFFs in canonical parametrization (\ref{Tpi}) are connected
with commonly used (see, e.g., \cite{PoS18}) by the following relations:
\begin{equation}
G^{(\pi)}_{10}(t) = A^{(\pi)}(t)\;,\quad G^{(\pi)}_{60}(t) = -\frac{1}{2}D^{(\pi)}(t)\;.
\label{GtoA}
\end{equation}

In the case of spin 1/2, Eq. (\ref{taufin}) gives the following result
which we will use below as the constituent-quark EMT canonical
parametrization:
$$
\langle p,m\left|T^{(q)}_{\mu\nu}(0)\right|p',m'\rangle =
\sum_{m''}\langle m\left|D_w^{1/2}(p,p')\right|m''\rangle\times
$$
$$
\langle m''\left|(1/2)g^{(q)}_{10}(t)K\,'_\mu K'_\nu
+ ig^{(q)}_{40}(t)\left[K'_\mu\,R_\nu + R_\mu\,K'_\nu\right] +\right.
$$
\begin{equation}
\left. + g^{(q)}_{60}(t)\left[tg_{\mu\nu} - K_\mu K_\nu\right]
\right|m'\rangle \;,
\label{Tq}
\end{equation}
These GFFs in the canonical parametrization (\ref{Tq}) can be written in
terms of commonly used GFFs for particles of spin 1/2 in the form
$$
g^{(q)}_{10}(t) = \frac{1}{\sqrt{1-t/4M^2}}\times
$$
\begin{equation}
\left[\left(1 - \frac{t}{4M^2}\right)
A^{(q)}(t)
+ 2\frac{t}{4M^2}J^{(q)}(t)\right]\;,
\label{g10}
\end{equation}
\begin{equation}
g^{(q)}_{40}(t) = -\,\frac{1}{M^2}\frac{J^{(q)}(t)}{\sqrt{1 - t/4M^2}})\;,
\label{g40}
\end{equation}
\begin{equation}
g^{(q)}_{60}(t) = -\,\frac{1}{2}\sqrt{1 - \frac{t}{4M^2}}D^{(q)}(t)\;.
\label{g60}
\end{equation}

In the following Section we generalize the method of construction of the
EMT matrix elements, given above, to composite systems.

\section{The EMT matrix elements for a system of two\\ free
particles with pion quantum numbers}
\label{sec: Sec 3}

Our relativistic approach to form factors of composite systems of
interacting components makes use of form factors of corresponding free
systems. So, to obtain GFFs of a composite system we need to construct GFFs
that describe the gravitational properties of a system of two free
constituents, the two-particle system as a whole having quantum
numbers of the composite system under consideration. We call the GFFs of
the two-particle system without interaction the free gravitational
two-particle form factors. The form factors of a composite system of two
interacting particles are written in our approach in terms of free
two-particle form factors and wave functions exploiting modified impulse
approximation (MIA). This approximation was first formulated in the case
of electroweak properties of hadrons in our papers
\cite{KrT02, KrT03} (see also the review \cite{KrT09}).

EMT operator $T^{(0)}_{\mu\nu}(0)$ for a system of two free particles is
of the form
\begin{equation}
T^{(0)}_{\mu\nu}(0) =
T_{1\,\mu\nu}\otimes I^{(2)}\oplus T_{2\,\mu\nu}\otimes I^{(1)}\;.
\label{T=T1IT2}
\end{equation}
Here $T_{1,2\,\mu\nu}$ are EMTs of the particles, and $I^{(1,2)}$ are
the identity operators in one-particle Hilbert-state spaces of the
particles.
The following set of two-particle vectors can be chosen as the basis:
\begin{equation}
|\,\vec p_1\,,m_1;\,\vec p_2\,,m_2\,\!\rangle  =
|\,\vec p_1\,m_1\,\!\rangle \otimes
|\, \vec p_2\,m_2\,\rangle\;,
\label{p1p2}
\end{equation}
where $\vec p_1,\,\vec p_2$ are the 3-momenta of particles,
$m_1,\,m_2$ are the projections of spins to the $z$ axis,
the normalization of one-particle vectors is given in (\ref{norm1}).
In terms of matrix elements in the basis
(\ref{p1p2}),  the relation (\ref{T=T1IT2})
is rewritten as the sum of matrix elements of one-particle EMT operators
$$
\langle\vec p_1,m_1;\vec p_2,m_2|T_{\mu\nu}^{(0)}(0)| \vec
p\,'_1,m'_1;\vec p\,'_2,m'_2\rangle =
$$
$$
=\langle\vec p_1,m_1|\vec
p\,'_1,m'_1\rangle \langle\vec p_2,m_2|T_{2\,\mu\nu}(0)|\vec
p\,'_2,m'_2\rangle +
$$
\begin{equation}
+ (1\leftrightarrow 2)\;.
\label{T=T1MT2}
\end{equation}
Each of the matrix elements of the one-particle EMT in
(\ref{T=T1MT2})
can be written in terms of GFFs (\ref{Tq})
(see, e.g., \cite{KrT02, KrT03, KoT72}).

Along with this basis (\ref{p1p2}), we consider the basis in which the
motion of the center  of mass of two particles is separated
(\cite{KrT02, KrT03, KoT72}):
$$
|\,\vec P,\;\sqrt {s},\;J,\;l,\;S,\;m_J\,\rangle\;,
$$
$$
\langle\,\vec P,\;\sqrt {s},\;J,\;l,\;S,\;m_J |\,\vec
P\,',\;\sqrt {s'},\;J',\;l',\;S',\;m_{J'}\,\rangle =
$$
$$
= N_{CG}\,\delta^{(3)}(\vec P - \vec P\,')\delta( \sqrt{s} -
\sqrt{s'})\times
$$
\begin{equation}
\delta_{JJ'}\delta_{ll'}\delta_{SS'}\delta_{m_Jm_{J'}}\;,
\label{Pk}
\end{equation}
$$
N_{CG} = \frac{(2P_0)^2}{8\,k\,\sqrt{s}}\;,\quad k =
\frac{\sqrt{\lambda(s\,,\,M^2\,,\,M^2)}}{2\,\sqrt{s}}\;,
$$
where $P_\mu = (p_1 +p_2)_\mu$, $P^2_\mu = s$, $\sqrt {s}$
is the invariant mass of the  system of two
particles, $l$ is the orbital momentum in the center-of-mass system
(c.m.s.), $\vec S\,^2=(\vec S_1 + \vec S_2)^2 = S(S+1)\;,\;S$ is the
total spin in c.m.s., $J$
is the total angular momentum, $m_J$ is the projection of the total
angular momentum, $M$ is the constituent mass, and
$\lambda(a,b,c) = a^2 + b^2 + c^2 - 2(ab + ac + bc)$.
The basis (\ref{Pk}) is related to the basis (\ref{p1p2})
by the Clebsch-Gordan decomposition
for the Poincar\'e group. The corresponding decomposition of a direct
product (\ref{p1p2})  of two irreducible representations of the Poincar\'e
group into irreducible representations (\ref{Pk})  for particles with spin
1/2 has the form \cite{KoT72} (see also \cite{KrT09}):
$$
|\,\vec
p_1\,,m_1;\,\vec p_2\,,m_2\,\rangle  = \sum |\,\vec P,\;\sqrt {s},
\;J,\;l,\;S,\;m_J\,\rangle
$$
$$
\times \langle Jm_J|S\,l\,m_S\,m_l\,\rangle Y^*_{lm_l}(\vartheta
\,,\varphi )\langle S\,m_S\,|1/2\,1/2\,\tilde m_1\,\tilde
m_2\,\rangle
$$
\begin{equation}
\times \langle\,\tilde
m_1|\,D_w^{1/2}(P,p_1)\,|m_1\,\rangle \langle\,\tilde
m_2\,|\,D_w^{1/2}(P,p_2)\,|m_2\,\rangle\;,
\label{KGpr}
\end{equation}
where $\vec p = (\vec p_1 - \vec p_2)/2$, $p = |\vec p|$,
$\vartheta \,,\varphi$ are the spherical angles of the vector $\vec p$ in
c.m.s., $Y_{lm_l}$ is the spherical function,
$\langle\,S\,m_S\,|1/2\,1/2\,\tilde m_1\,\tilde m_2\,\rangle $ and
$\langle Jm_J|S\,l\,m_S\,m_l\,\rangle $ are the Clebsh-Gordan coefficients
of the group $SU(2)$, $\langle\,\tilde m|\,D_w^{1/2}(P,p)\,|m\,\rangle $
is the matrix of the
three-dimensional spin rotation, that is necessary for the
relativistic  invariant summation of the particle spins. The sums go
over all discrete variables ,
$\tilde m_1$, $\tilde m_2$, $m_l$, $m_S$, $l$, $S$, $J$, $m_J$.
To obtain the basis where the center-of-mass motion is separated we invert
the decomposition (\ref{KGpr}):
$$
|\,\vec P,\;\sqrt {s},\;J,\;l,\;S,\;m_J\,\rangle =
\sum_{m_1\;m_2}\,\int \,\frac {d\vec p_1}{2p_{10}}\, \frac {d\vec
p_2}{2p_{20}}\times
$$
$$
|\,\vec p_1\,,m_1;\,\vec p_2\,,m_2\,\rangle \times
$$
\begin{equation}
\times \langle\,\vec p_1\,,m_1;\,\vec p_2\,,m_2\,| \,\vec
P,\;\sqrt {s},\;J,\;l,\;S,\;m_J\,\rangle\;,
\label{Klebsh}
\end{equation}
with the Clebsh-Gordan coefficient
$$
\langle\,\vec p_1\,,m_1;\,\vec p_2\,,m_2\,|\,\vec P,\;\sqrt {s},
\;J,\;l,\;S,\;m_J\,\rangle =
$$
$$
=\sqrt {2s}[\lambda
(s,\,M^2,\,M^2)]^{-1/2}\,2P_0\,\delta (P - p_1 - p_2)\times
$$
$$
\times \sum \langle\,m_1|\,D_w^{1/2}(p_1\,,P)\,|\tilde m_1\,\rangle
\langle\,m_2|\,D_w^{1/2}(p_2\,,P)\,|\tilde m_2\,\rangle
$$
$$
\times \langle{1}/{2}\,{1}/{2}\,\tilde m_1\,\tilde
m_2\,|S\,m_S\,\rangle \,Y_{lm_l}(\vartheta\,,\varphi )\, \langle
S\,l\,m_S\,m_l\,|Jm_J\rangle\;,
$$
the sum being over $\tilde m_1,\,\tilde m_2,\,m_l,\,m_S$.

We use below the basis (\ref{Klebsh}) with pion quantum numbers
$J=l=S=0$:
\begin{equation}
|\,\vec P,\;\sqrt {s},\;0,\;0,\;0,\;0\,\rangle = |\,\vec P,\;\sqrt {s}\,\rangle\;.
\label{Ps0000}
\end{equation}
We construct the EMT matrix element in the basis (\ref{Klebsh}) for
quantum numbers given above using the general method of parametrization of
Section \ref{sec: Sec 2}. Using (\ref{T=tau})-(\ref{Gi}),
(\ref{taufin}) we obtain the parametrization which is analogous to
that for zero spin (\ref{Tpi}):
$$
\langle P,\sqrt{s}\left|T^{(0)}_{\mu\nu}(0)\right|P',\sqrt{s'}\rangle =
$$
$$
= \frac{1}{2}G^{(0)}_{10}(s,t,s')A'_\mu A'_\nu +
$$
\begin{equation}
+ G^{(0)}_{60}(s,t,s')\left[t\,g_{\mu\nu} - A_\mu A_\nu\right]\;,
\label{T0}
\end{equation}
where $G^{(0)}_{i0}(s,t,s'),\,i=1,6$ are free two-particle GFFs,
$$
A_\mu = \left(P - P'\right)_\mu\;,\quad A^2 = t\;,
$$
$$
A'_\mu = \frac{1}{(-t)}\left[(s - s' - t)P_\mu + (s' - s -
t)P'_\mu\right]\;.
$$
It is easy to show that all the imposed constraints are satisfied.

It is possible to derive the equations analogous to (\ref{T0})
for free two-particle systems with different quantum numbers.
Such constructions were obtained in
\cite{KrP16, KrT02, KrT03, KoT72}
in the context of the parametrization of the matrix elements of
electroweak currents.

Note that the objects $G^{(0)}_{i0}(s,t,s'),\,i=1,6$, in general, are
generalized functions (distributions), defined on a space of test functions
(see, e.g., \cite{BoL90}, and also \cite{LoC17, KrT02,
KrT03, KrT05}), and so the static limits at $t\to 0$ are to be
understood in a weak sense.
The functionals generated by free two-particle form factors on
the space of the two-quark wave functions of pion give the
corresponding pion GFFs in MIA (see Sect. \ref{sec: Sec 4} below).

The free two-particle form factors in (\ref{T0}) can be written in terms
of one-particle GFFs realizing the parametrization of the matrix elements
(\ref{T=T1MT2}), namely, in terms of the form factors of
constituent quarks (\ref{Tq}),(\ref{g10})-(\ref{g60}). Using the
decomposition (\ref{Klebsh}), we obtain the matrix element (\ref{T0}) in
the following form,
$$
\langle P,\sqrt{s}\left|T^{(0)}_{\mu\nu}(0)\right|P',\sqrt{s'}\rangle =
$$
$$
= \sum\int\frac{d\vec p_1}{2p_{10}}\frac{d\vec p_2}{2p_{20}}\frac{d\vec p\,'_1}{2p'_{10}}\frac{d\vec p\,'_2}{2p'_{20}}
\langle P,\sqrt{s}\left|\right.\vec p_1,m_1;\vec p_2,m_2\rangle\times
$$
$$
\left[\langle \vec p_1,m_1\left|\right.\vec p\,'_1,m'_1\rangle \langle p_2,m_2\left|T^{(2)}_{\mu\nu}(0)\right|p'_2,m'_2\rangle\right. +
$$
$$
+ \left.\langle \vec p_2,m_2\left|\right.\vec p\,'_2,m'_2\rangle \langle p_1,m_1\left|T^{(1)}_{\mu\nu}(0)\right|p'_1,m'_1\rangle\right]\times
$$
\begin{equation}
\langle \vec p\,'_1,m'_1;\vec p\,'_2,m'_2\left|\right.
P',\sqrt{s'}\rangle\;,
\label{T0p1p2}
\end{equation}
where the sums are over the variables $m_1,\,m_2,\;m'_1,\,m'_2$.

The substitution of (\ref{T0}), one-particle matrix elements of EMT
(\ref{Tq}), and the Clebsh-Gordan coefficiemts (\ref{Klebsh}) for
$J=l=S=0$ in (\ref{T0p1p2}) gives the desired free two-particle GFFs.
The integrals in (\ref{T0p1p2}) are written in the coordinate frame with
$\vec P\,'=0,\;\vec P=(0,0,P)$. The $D_w$-functions for spin 1/2 are of
the form \cite{Che66}:
$$
D_w^{1/2}(p_1,p_2) = \cos (\omega /2) - 2i(\vec k\hat{\vec j})\,\sin
(\omega /2)\;,
$$
$$
\vec k = \frac {[\,\vec
p_1\,\vec p_2\,]}{|[\,\vec p_1\,\vec p_2\,]|}\;,
$$
\begin{equation}
\omega =
2\,\arctan\frac {|[\,\vec p_1\,\vec p_2\,]|}{(p_{10} +
M_1)(p_{20} +M_2) - (\vec p_1\vec p_2)}\;,
\label{D1/2}
\end{equation}
where $\hat{\vec j}$ is the operator of the particle spin written in
terms of the Pauli matrices. In the choosen coordinate system, two
$D_w$-functions in r.h.s. of (\ref{T0p1p2}) become unity
matrices and the other are written with the use of (\ref{D1/2}). The
sum of the rotations around the same axis is obtained following the
prescription:
\begin{equation}
D_w^{1/2}(\omega_1)D_w^{1/2}(\omega_2) = D_w^{1/2}(\omega_1+\omega_2)\;.
\label{Dsum}
\end{equation}
After performing the convolution of both sides, first, with the tensor
$A'\,^\mu A'\,^\nu$, second, with $g^{\mu\nu}$, and the integrations
and summations, we obtain the system of two algebraic equations for the
free form factors $G^{(0)}_{i0}(s,t,s'),\,i=1,6$:
$$
\frac{1}{2}G^{(0)}_{10}\left[\frac{\lambda(s,t,s')}{t}\right]^2 -
\lambda(s,t,s')G^{(0)}_{60}=
$$
$$
= \frac{1}{2}A\left\{\frac{1}{2}\left[g^{(u)}_{10}(t)+g^{(\bar d)}_{10}(t)\right]
(s + s'-t)^2\cos(\omega_1+\omega_2) - \right.
$$
$$
- M\left[g^{(u)}_{40}(t)+g^{(\bar d)}_{40}(t)\right]\xi(s,t,s')(s+s'-t)\sin(\omega_1+\omega_2)-
$$
\begin{equation}
- \left.\left[g^{(u)}_{60}(t)+g^{(\bar d)}_{60}(t)\right]\lambda(s,t,s')\cos(\omega_1+\omega_2)\right\}\;,
\label{eq1}
\end{equation}
$$
\frac{1}{2}G^{(0)}_{10}\left[\frac{\lambda(s,t,s')}{(-t)}\right] +
3\,t\,G^{(0)}_{60}=
$$
$$
= \frac{1}{2}A\left\{\frac{1}{2}\left[g^{(u)}_{10}(t)+g^{(\bar d)}_{10}(t)\right]
(4M^2- t)\,\cos(\omega_1+\omega_2) + \right.
$$
\begin{equation}
+ 3\,t\left.\left[g^{(u)}_{60}(t)+g^{(\bar d)}_{60}(t)\right]\cos(\omega_1+\omega_2)\right\}\;,
\label{eq2}
\end{equation}
where $A=A(s,t,s')=2R(s,t,s')\lambda(s,t,s'),$
$$
R(s, t, s') = \frac{(s + s'- t)}{2\sqrt{(s-4M^2) (s'-4M^2)}}\,
$$
$$
\times\frac{\vartheta(s,t,s')}{{[\lambda(s,t,s')]}^{3/2}}\;,
$$
$$
\xi(s,t,s')=\sqrt{-(M^2\lambda(s,t,s')+ss't)}\;,
$$
$\omega_1$ and $\omega_2$ are the Wigner spin-rotation parameters:
$$
\omega_1 =
\arctan\frac{\xi(s,t,s')}{M\left[(\sqrt{s}+\sqrt{s'})^2 -
t\right] + \sqrt{ss'}(\sqrt{s} +\sqrt{s'})}\;,
$$
$$
\omega_2 = \arctan\frac{ \alpha (s,s') \xi(s,t,s')} {M(s + s' -
t) \alpha (s,s') + \sqrt{ss'}(4M^2 - t)}\;,
$$
$\alpha (s,s') = 2M + \sqrt{s} + \sqrt{s'} $,
$\vartheta(s,t,s')= \theta(s'-s_1)-\theta(s'-s_2)$, $\theta$ is the
Heaviside function.
$$
s_{1,2}=2M^2+\frac{1}{2M^2} (2M^2-t)(s-2M^2)
$$
$$
\mp \frac{1}{2M^2} \sqrt{(-t)(4M^2-t)s(s-4M^2)}\;,
$$
$g^{(u,\bar d)}_{i0}(t)\;,\;i=1,4,6$ the GFFs of  $u$- and
$\bar d$- quarks, respectively.
The cutting off by the Heaviside functions in (\ref{eq1}), (\ref{eq2})
gives the kinematically available region in the plane of invariant
variables $(s,s')$ (see, e.g., \cite{KrT09}).

The formal solution of the system (\ref{eq1}), (\ref{eq2}) is of the
form:
$$
G^{(0)}_{10}(s, t, s') = \frac{R(s, t, s')\,t}{\lambda(s,t,s')}
$$
$$
\times\left\{\frac{1}{2}\left[g^{(u)}_{10}(t)+g^{(\bar d)}_{10}(t)\right]\left[(4M^2-t)\lambda(s,t,s')\right.\right. +
$$
$$
+ \left. 3\,t\,(s + s'-t)^2\right]\cos(\omega_1+\omega_2) -
$$
$$
- 3Mt\left[g^{(u)}_{40}(t)+g^{(\bar d)}_{40}(t)\right]\times
$$
\begin{equation}
\left.\xi(s,t,s')(s+s'-t)\sin(\omega_1+\omega_2)\right\}\;,
\label{G10}
\end{equation}
$$
G^{(0)}_{60}(s, t, s') = \frac{1}{2}\,R(s, t, s')
$$
$$
\times\left\{\frac{1}{2}\left[g^{(u)}_{10}(t)+g^{(\bar d)}_{10}(t)\right]
\left[(s + s'- t)^2 +\right.\right.
$$
$$
+ \left.(4M^2-t)\lambda(s,t,s')/t\right]
\cos(\omega_1+\omega_2) -
$$
$$
- M\left[g^{(u)}_{40}(t)+g^{(\bar d)}_{40}(t)\right]\times
$$
$$
\left.\xi(s,t,s')(s+s'-t)\sin(\omega_1+\omega_2)\right. +
$$
\begin{equation}
+ \left. 2\left[g^{(u)}_{60}(t)+g^{(\bar d)}_{60}(t)\right]\lambda(s,t,s')\cos(\omega_1+\omega_2)\right\}\;.
\label{G60}
\end{equation}
Note that the system (\ref{eq1}), (\ref{eq2}) is ill-defined at $t\to 0$:
the corresponding determinant is zero for $t=0$. So, the solution
for the form factor (\ref{G60}) does not exist at $t=0$, and the weak limit
at $t\to 0$ of the form factor (\ref{G60}), considered as a regular
generalized function on the space of the test functions, is infinite. The
singularity $\sim 1/t$ is contained in the term with quark form factors
$g^{(u)}_{10}(t)+g^{(\bar d)}_{10}(t)$.

Let us argue that the occurrence of this singularity does not discard
the approach but rather puts it into the general trend. To clarify the
physical meaning of the singularity in the case of the free two-particle
system, we consider the expression for the mean-square
mechanical radius \cite{PoS18} in the following form:
\begin{equation}
\langle r^2\rangle_{\rm mech} = \lim\limits_{R\to\infty}
\frac{\int\limits_0^R\,d^3r\,r^2\left(\frac{2}{3}s(r) + p(r)\right)}
{\int\limits_0^R\,d^3r\,\left(\frac{2}{3}s(r) + p(r)\right)}\;,
\label{r2mmr}
\end{equation}
 where $s(r)$ and $p(r)$ are the longitudinal and transverse mechanical
stresses in a system, correspondingly.

The fact that the form factors (\ref{G10}), (\ref{G60}) describe
the properties of a system of two point-like particles without interaction
between them, means that there are no mechanical stresses in the
system. If we let $s(r)$ and $p(r)$ in (\ref{r2mmr}) be constant, and then
let these constants vanish, we would obtain that the mechnical MSR in the
system is infinite. On the other hand, this MSR is of the form
\cite{PoS18},
\begin{equation}
\langle r^2\rangle_{\rm mech} =
\frac{6}{D(t)}\,\left.\frac{dD}{dt}\right|_{t=0}\;,
\label{r2mD}
\end{equation}
where $D(t)$ is a functional generated by a regular distribution
(\ref{G60}) on the suitable space of test functions.

The equation (\ref{r2mD}), in analogy to
(\ref{r2mmr}), gives the infinity for the value  of the mechanical MSR
if the functional $D(t)$, and, consequently, the distribution (\ref{G60}),
are singular at the point $t=0$. So, we conclude that it is
adequate to use the free two-particle form factor with singularity
(\ref{G60}) for calculations.

Now we include the interaction in this system and derive the pion
$D$ form factor using MIA (see
Sect. \ref{sec: Sec 4}). To obtain a finite mechanical MSR of the pion,
we need to regularize
(\ref{G60}) in the vicinity of the point $t=0$, that is to find a
closely related non-singular function which gives similar physical
results. Here we choose to appeal to
the nonrelativistic case.
For the
simpliest variant of our relativistic composite model that we use in the
present paper it is sufficient to present an {\it ansatz} for constructing
the free two-particle form factor in a small neighbourhood of $t=0$ and to
show that the construction has a narrow range of choice.

The main point is the fact that in the nonrelativistic limit of
(\ref{G60}) the first two terms vanish. So, the limit does not contain
the singularity and is defined by the third term with quark form factors
$g^{(u)}_{60}(t)+g^{(\bar d)}_{60}(t)$, having the following form,
\begin{equation}
G^{(0)}_{60nr}(k,t,k') = 2\left[g^{(u)}_{60}(t)+g^{(\bar d)}_{60}(t)\right]
\frac{\vartheta(k,t,k')}{k\,k'\,\sqrt{(-t)}}\;,
\label{G060n}
\end{equation}
$$
\vartheta(k,t,k')= \theta\left(k'-\left|k-\sqrt{(-t)}/2\right|\right) -
$$
$$
\theta\left(k'-\left(k+\sqrt{(-t)}/2\right)\right)\;.
$$
The quantity $G^{(0)}_{60nr}(k,t,k')$ is, in fact, a free nonrelativistic
two-particle form factor, the nonrelativistic analog of the form factor
(\ref{T0}), (\ref{G60}). The weak limit of the form factor (\ref{G060n}) at
$t\to 0$ is finite.

We require, firstly, that the nonrelativistic limit of the
regularized construction for the free
two-particle form factor near $t=0$ coincides with (\ref{G060n}). Further,
we take into account the fact that, usually, nonrelativistic models give
reasonable results at low momentum transfer. So, we require, secondly, that
our relativistic {\it ansatz} gives in MIA for pion at low $t$ the results
close to nonrelativistic results. This requirement means, in particular,
that the assumed  construction depends on the quark form factors
$g^{(u)}_{60}(t)+g^{(\bar d)}_{60}(t)$ only. In this case, if
$g^{(u)}_{60}(t)+g^{(\bar d)}_{60}(t)\to 0$, then the construction, as well
as the nonrelativistic expression (\ref{G060n}), goes to zero. Both these
requirements are satisfied by the third term in (\ref{G060n}). So, we
choose our {\it ansatz} in the form,
$$
G^{(0)}_{60}(s,t,s') = G^{(0a)}_{60}(s,t,s') = R(s,t,s')\lambda(s,t,s')\times
$$
\begin{equation}
\left[g^{(u)}_{60}(t)+g^{(\bar d)}_{60}(t)\right]\cos(\omega_1+\omega_2)\;,
\label{G60anz}
\end{equation}
where $G^{(0a)}$ denotes the functions that appear as a result of the
supposition.

It is clear that even for the chosen assumption, based on the
nonrelativistic limit, the actual proposed form  is not unique. It is
possible to add to (\ref{G60anz}) some arbitrary functions which do not
change its nonrelativistic limit (\ref{G060n}). However, the second
condition requires that the contribution of these functions near $t=0$ is
small: the results have to be close to the nonrelativistic case. This
means that qualitatively the added terms must not change the result. So,
in what follows we use (\ref{G60}) for the pion $D$-formfactor at finite
values of $t$, but in the vicinity of the point $t=0$ we make use of
(\ref{G60anz}).

\section{Gravitational form factors of pion in modified impulse approximation}
\label{sec: Sec 4}

To obtain the GFFs of pion we use instant form (IF) of RQM
(\cite{Dir49, LeS78, KeP91, Coe92}). The details of our version for
composite systems can be foud in the review
(\cite{KrT09}).
In RQM the interaction operator is included in the generators of the
Poincar\'e group, the commutation relations of the algebra being preserved.
We include the interaction in the algebra of the Poincar\'e group
following the procedure of \cite{BaT53}:
\begin{equation}
\hat M_0
\to \hat M_I = \hat M_0 + \hat V \;,
\label{M0toMI}
\end{equation}
here $\hat M_0$ is the operator of the invariant mass for a free system,
$\hat V$ is interaction operator, and $\hat M_I$ the mass operator for the
system with interaction.

The wave function of the system of interating particles in IF RQM is
defined as the eigenfunction of the following complete set of the
operators:
\begin{equation}
{\hat M}_I^2\;(\hbox{or}\;\hat M_I)\;,\quad
{\hat J}^2\;,\quad \hat J_3\;,\quad \hat {\vec P}\;,
\label{complete}
\end{equation}
here ${\hat J}^2$ is the operator of the square of the total angular
moment, $\hat J_3$ is the operator of the projection of the total
angular moment on the $z$ axis and $\hat {\vec P}$ is the operator of the
total momentum.

In the IF RQM the operators ${\hat J}^2\;,\; \hat J_3\;,\; \hat {\vec P}$
coincide with corresponding operators for the composite system without
interaction and only the term $\hat M_I^2\;(\hat M_I) $
is interaction depending.
The two-quark wave function of pion in the basis given by the
complete set of vectors (\ref{Pk}), (\ref{Klebsh}), (\ref{Ps0000})
diagonalizes (\ref{complete}) and has the form:
$$
\langle\vec P\,,\,\sqrt {s}|\vec p\rangle  =
N_C\,\delta (\vec P\, - \vec p)\,\varphi(k)\;,
$$
\begin{equation}
N_C =
\sqrt{2p_0}\sqrt{\frac{N_{CG}}{4\,k}}\;,
\label{wfI}
\end{equation}
The wave function of intrinsic motion is the eigenfunction of the operator
$\hat M_I^2\;(\hat M_I)$ and in the case of two particles of equal masses
is
\begin{equation}
\varphi(k(s)) =\sqrt[4]{s}\,u(k)\,k\;,\quad
\int\,u^2(k)\,k^2\,dk = 1\;,
\label{phi(s)}
\end{equation}
The normalization factors in (\ref{phi(s)}) correspond to the transition
to the relativistic density of states
\begin{equation}
k^2\,dk\quad \to\quad \frac{k^2\,dk}{2\sqrt{k^2 + M^2}}\;.
\label{rel den}
\end{equation}

The decomposition of the matrix element (\ref{Tpi}) of the pion EMT in
terms of the complete set of the vectors  (\ref{Pk}), (\ref{Klebsh}),
(\ref{Ps0000}) is
$$
\langle \vec p\,\left|T^{(\pi)}_{\mu\nu}(0)\right|\vec p\,'\rangle =
$$
$$
= \int\,\frac{d\vec P\,d\vec
P\,'}{N_{CG}\,N'_{CG}}\,d\sqrt{s}\,d\sqrt{s'}\, \langle \vec p\,|\vec
P\,,\sqrt{s}\,\rangle\times
$$
\begin{equation}
\langle\vec P\,,\sqrt{s}\,|T^{(\pi)}_{\mu\nu}(0)|\vec
P\,'\,,\sqrt{s'}\rangle\langle\vec P\,'\,,\sqrt{s'}\,|\vec p\,'\rangle
\;,
\label{int=T}
\end{equation}
where $\langle\vec P\,'\,,\sqrt{s'}|\vec p\,'\rangle$ is the wave
function in the sense of IF RQM (\ref{wfI}).We obtain
$$
\langle \vec p\,\left|T^{(\pi)}_{\mu\nu}(0)\right|\vec p\,'\rangle =
$$
$$
=\int\,\frac{N_C\,N'_C}{N_{CG}\,N_{CG}'}\,d\sqrt{s}\,d\sqrt{s'}\,
\varphi(s)\times
$$
\begin{equation}
\langle\vec p\,,\sqrt{s}\left|T^{(\pi)}_{\mu\nu}(0)\right|\vec
p\,'\,,\sqrt{s'}\rangle \varphi(s')\;.
\label{int ds=T}
\end{equation}
The matrix element of the tensor in (\ref{int ds=T}) is to be considered
as a Lorentz-covariant generalized function
\cite{BoL90, KrT02, KrT03, KrT05}), that has a meaning only under the
integral. The integral itself presents a functional giving a regular
distribution.
The decomposition of the tensor in the integral in terms of tensors which
were used in the decomposition in l.h.s. of (\ref{int ds=T}) entering
 (\ref{Tpi}) is:
$$
\frac{N_C\,N'_C}{N_{CG}\,N'_{CG}}\langle\vec p\,,\sqrt{s}\left|T^{(\pi)}_{\mu\nu}(0)\right|\vec
p\,'\,,\sqrt{s'}\rangle \varphi(s') =
$$
$$
= \frac{1}{2}\tilde G_{10}(s,t,s')K'_\mu K'_\nu +
$$
\begin{equation}
+ \tilde G_{60}(s,t,s')\left[tg_{\mu\nu} - K_\mu K_\nu\right]\;,
\label{Tgenf}
\end{equation}
here $\tilde G_{i0}(s,t,s'),\,i=1,6$ are the
Lorentz-invariant regular distributions.
A rigorous proof of the importance of distributions in the interpretation
of the decomposition analogous to (\ref{Tgenf}) in the case of
electromagnetic  current was given in \cite{KrT02, KrT03}.
After substituting of (\ref{Tpi}) and (\ref{Tgenf}) in l.h.s. and r.h.s. of
(\ref{int ds=T}), respectively, we obtain pion GFFs in the form of
functionals:
$$
G^{(\pi)}_{i0}(t) =
\int\,d\sqrt{s}\,d\sqrt{s'}\,
\varphi(s)\tilde G_{i0}(s,t,s')\varphi(s')\;,
$$
\begin{equation}
i=1,6\;.
\label{int ds=Gpi}
\end{equation}

The main point now is the calculation of the function $\tilde G_{i0}(s,t,s')$.
To obtain similar form factors describing electroweak structure of
composite hadrons it is customary exploit the so-called impulse
approximation (IA) (see, e.g., the review \cite{KeP91}). Let us
demonstrate the meaning of IA extending the approach  to GFF. These form
factors characterize the scattering cross-section of a projectile by a
composite system in the process of graviton exchange.

So, the EMT in this case can be written in the following form:
\begin{equation}
T = \sum_{k}\,T^{(k)} +
\sum_{k\langle m}\,T^{(km)}+\ldots\;,
\label{T=Tk+Tkm}
\end{equation}
where the first term presents the sum of one-particle EMTs, the second
term presents the sum of two-particle EMT, and so on. The first sum
describes the scattering of a projectile by each independent
constituent, the second sum describes the scattering by two constituents
simultaneously and so on.
The standard IA leaves in (\ref{T=Tk+Tkm}) only the first term:
\begin{equation}
T \approx \sum_{k}\,T^{(k)}\;.
\label{IA}
\end{equation}
Note that in the approximation (\ref{IA}) the operators
in the instant form RQM does not satisfy the
Lorentz-covariance conditions and the conservation law
~\cite{KeP91}.

To study the electroweak structure of hadrons, we had proposed
\cite{KrT02, KrT03} the modified impulse approxination
(MIA). Constructing MIA for GFFs we change the form factors
$\tilde G_{i0}(s,t,s')$ in (\ref{int ds=Gpi}) for free two-particle GFFs
(\ref{T0}): in the invariant part of the decomposition
(\ref{int ds=T}), (\ref{Tgenf}) we throw off the contribution of the
simultaneous scattering by two and more constituents and take into
account only scattering by free two-constituent system.

The covariant part of the decomposition  (\ref{int ds=T}) -- (\ref{int
ds=Gpi}) is not changed by MIA and so,
the Lorentz-covariance conditions and the conservation law
for the EMT matrix element (\ref{int ds=T}) are not broken.
This happens because in MIA the contribution of the second term in
(\ref{T=Tk+Tkm}) is partially taken into account in a self-consistent way.

In MIA, the pion GFFs (\ref{int ds=Gpi}) are written in the
form:
$$
G^{(\pi)}_{i0}(t) =
\int\,d\sqrt{s}\,d\sqrt{s'}\,
\varphi(s)\,G^{(0)}_{i0}(s,t,s')\varphi(s')\;,
$$
\begin{equation}
i=1,6\;,
\label{GpiMIA}
\end{equation}
where $G^{(0)}_{i0}(s,t,s')$ are free two-particle form factors
(\ref{T0}), given by (\ref{G10}), (\ref{G60}),
(\ref{G60anz}).

In the following Section, the details of calculation of pion GFFs using
(\ref{GpiMIA}) and the corresponding results are given.

\section{Results of calculatuons}
\label{sec: Sec 5}

In what follows we use the conventional notations of
$A$, $J$ and $D$ form factors (see, e.g., \cite{PoS18}) and the linking
relations (\ref{GtoA}), (\ref{g10}) -- (\ref{g60}).

To obtain the pion form factor $A$ we use directly the equations
(\ref{GtoA}), (\ref{GpiMIA}), (\ref{G10}) while in the case of the form
factor $D$, which is ill-defined, we involve an ${\it anzatz}$ described in
detail in Section \ref{sec: Sec 3}.
We obtain the pion $D$ form factor in the vicinity of $t=0$
using (\ref{GtoA}), (\ref{GpiMIA}) and assuming (\ref{G60anz})
(the form factor $D^{(\pi a)}(t)$).
For the overall description of the pion $D$ form factor we need to join
smoothly this function $D^{(\pi a)}(t)$ with the solution for finite
values of $t$ given by (\ref{GtoA}), (\ref{GpiMIA}), (\ref{G60}).
We describe this procedure in detail later.

Let us list first the relativistic effects contained in
(\ref{GpiMIA}). The contributions of the  $J$ form factors of the
constituent quarks ($g^{(q)}_{40}(t)$) to the pion $A$
($G^{(\pi)}_{10}(t)$) and $D$ ($G^{(\pi)}_{60}(t)$) form factors are a
consequence of pure relativistic effect of spin rotation. These
contributions vanish if we set
$\omega_{1,2}$ in (\ref{G10}), (\ref{G60}) equal to zero.
The contribution of quark $A$ form factor $g^{(q)}_{10}(t)$ to pion
$D$ form factor (\ref{G60}) is of relativistic origin, too.

To obtain numerical results for pion GFFs in our model
(\ref{GpiMIA}) (\ref{G10}), (\ref{G60}),
we need some parameters to be used as an input.
We suppose that $u$- and $d$- quarks have one and the same gravitational
structure and so, we have to set three quark GFFs as functions of momentum
transfer square. It is also necessary to choose a model two-quark wave
function of pion (\ref{phi(s)}), and to fix the mass of light quark, $M$.

In the present work we consider the simpliest case, that of point-like
constituent quarks. This means that instead of quark form factors, we use
their standard static moments:
$$
A^{(q)}(t) = A^{(q)}(0) = 1\;,\quad J^{(q)}(t) = J^{(q)}(0) =\frac{1}{2}\;,
$$
\begin{equation}
D^{(q)}(t) = D^{(q)}(0) = D_q\;,\quad q = u, \bar d\;,
\label{AqA0}
\end{equation}
where $D_q$ is the $D$-term of the constituent quark.

We had shown \cite{KrT01} that the results of calculations for
 electromagnetic form factors depend weakly on the actual form of the
 two-quark wave function in pion. Here we choose for (\ref{phi(s)}) the
 wave function of the ground state of harmonic oscillator which ensures
 square-law quark confinement,
\begin{equation}
u(k) = \left(\frac{4}{\sqrt{\pi}\,b^3}\right)^{1/2}\exp\left(-\,\frac{k^2}{2\,b^2}\right)\;.
\label{wfHO}
\end{equation}
Here $b$ is the parameter of the model along with the quark-mass $M$.
The best results for electroweak properties of
light mesons \cite{KrT01, KrT98, KrT09prc, TrT13, KrP16, KrP18, KrT19,
KrT17, KrN13} were obtained for the following values of these parameters:
\begin{equation}
M = 0.22\;\mbox{GeV}\;,\quad b = 0.35\;\mbox{GeV}\;.
\label{Mb}
\end{equation}
In what follows we fix these values also for the calculation of pion GFFs.
So, to derive the pion GFFs we need to fix only the constituent-quark
$D$-term (\ref{AqA0}).

The mean-square radius (MSR) of pion we define as follows (see
\cite{PoS18} and the original paper \cite{ChS63}):
\begin{equation}
\langle r^2_\pi\rangle = 6\,A^{(\pi)}\,'(0) - \frac{3}{2\,M^2_\pi}D^{(\pi)}(0)\;,
\label{r2}
\end{equation}
where $M_\pi = 0.13957\,$GeV is the pion mass. Note that the standard
condition $A^{(\pi)}(0)=1$ is fulfilled automatically.

For the interval of possible data for the pion MSR we adopt the interval
that can be calculated using the results listed in the review
\cite{PoS18}, namely:
$$
\langle r^2_\pi\rangle_{min} = 65.38\;\mbox{GeV}^{-2}\;,
$$
\begin{equation}
\langle r^2_\pi\rangle_{max} = 69.52\;\mbox{GeV}^{-2}\;.
\label{rmm}
\end{equation}
To obtain this interval of MSR values in our approach, we require in
addition the parameter $D_q$ (\ref{AqA0}) to be in the following region
of approximately the same relative spread
\begin{equation}
D_q = -0.1435\pm 0.0045\;.
\label{DDq}
\end{equation}
The interval of values of the pion $D$-term corresponding to the chosen
interval of the quark $D$-term (\ref{DDq}) is
\begin{equation}
D^{(\pi)}(0)_{min} = -0.905\;,\quad
D^{(\pi)}(0)_{max} = -0.851\;.
\label{Dmm}
\end{equation}

The equations for the form factor $A^{(\pi)}(t)$ do not contain the
parameter $D_{q}$. So, the derivative of the $A$ form factor of pion at
$t=0$ is defined by the parameters (\ref{Mb}) fixed in our model approach
to the pion electroweak form factors and has a predictive nature. This
value is obtained numerically using (\ref{GtoA}), (\ref{GpiMIA}):
\begin{equation}
A^{(\pi)}\,'(0) = 0.0408\;\mbox{GeV}^{-2}\;.
\label{A0s}
\end{equation}

\begin{figure}[h!]
\epsfxsize=0.9\textwidth
\centerline{\psfig{figure=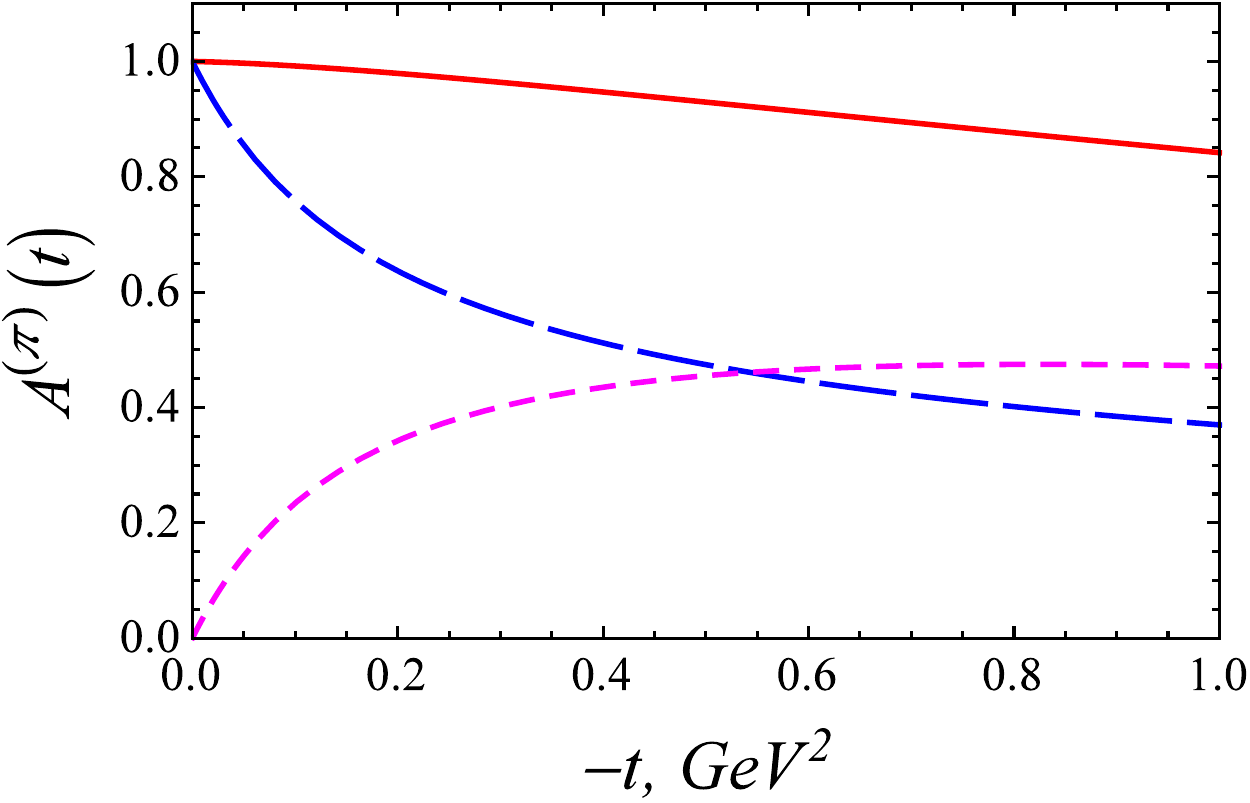,width=9cm}}
\vspace{0.3cm}
\caption{Gravitational $A$ form factor of pion. Full line (red) --
the full result; dashed line (blue) -- the contribution of the $A$ form
factors $g^{(q)}_{10}$ of the constituent quarks; short-dashed line
(magenta) -- the contribution of the quark $J$ form factors $g^{(q)}_{40}$
(relativistic spin rotation effect).}
\label{fig:1}
\end{figure}

\begin{figure}[h!]
\epsfxsize=0.9\textwidth
\centerline{\psfig{figure=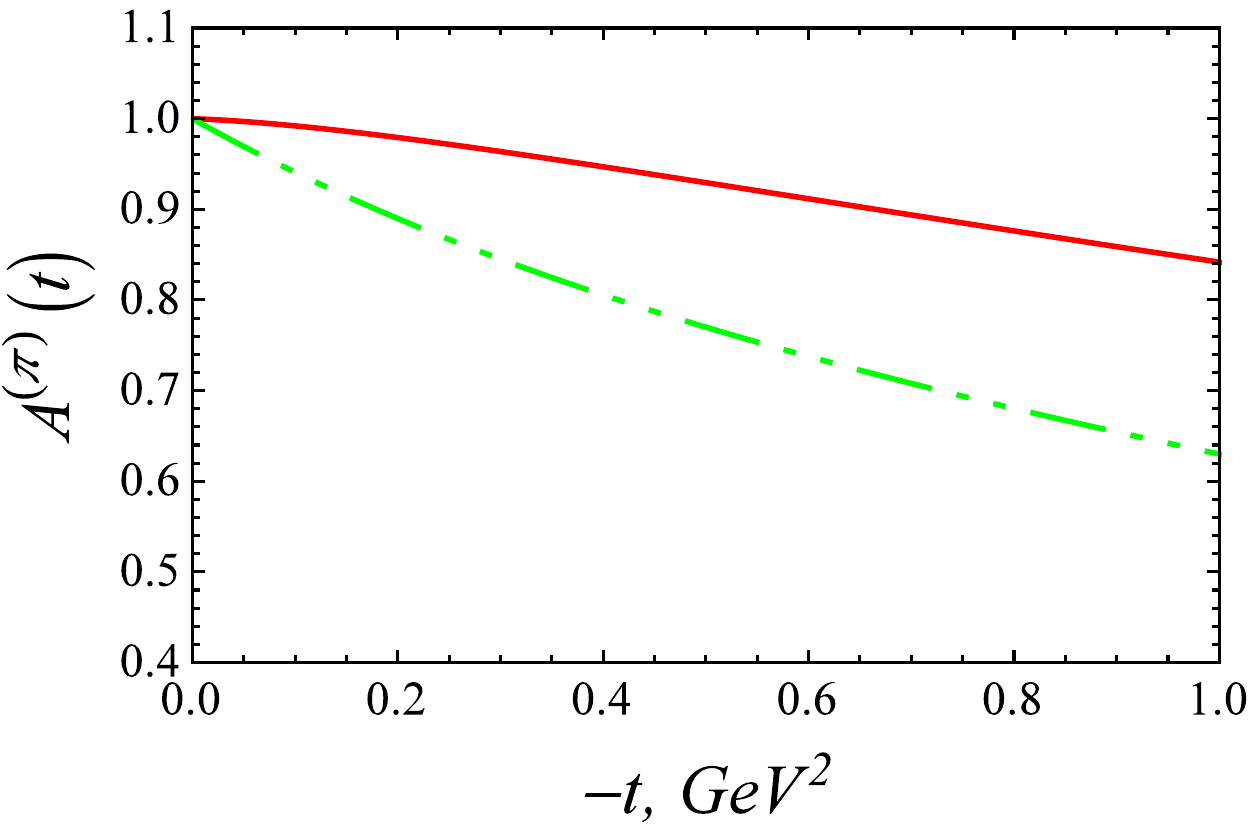,width=9cm}}
\vspace{0.3cm}
\caption{Our $A$ form factor of pion  (full line (red)) in
comparison with the result of the authors of \cite{KuS18}. Their $A$ form
factor of pion is normalized to its value at zero $t$ (double-dot-dashed,
green) line.}
\label{fig:2}
\end{figure}

The results of calculation of the pion $A$ form factor are presented in
fig. \ref{fig:1} and fig. \ref{fig:2}. Note that fig. \ref{fig:1}
demonstrates, in particular, that the relativistic spin rotation effect
gives an essential contribution to $A$ form factor. This effect
is purely kinematical and thus takes place for any model wave function.
The effect changes essentially the slope of $A$ form factor at zero $t$
and, as a consequence, the value of the pion gravitational radius
(\ref{r2}), (\ref{A0s}). This fact emphasizes the importance of the
corresponding theory to be essentially relativistic.

As we have mentioned above
recently the data on the pion GFFs was extracted  from the experiment
\cite{Mas16} for the first time in \cite{KuS18}. In fig. \ref{fig:2} we
compare our results for pion $A$ form factor with those given in
\cite{KuS18}. The results are in a qualitative agreement, however the
 slope of our $A$ form factor is smaller. Note that we choose here the
simpliest variant of the model confining ourseves to point-like
constituent quarks. If we depart from this condition, the quark form
factors would give the additional decreasing of the pion form factor and
would ameliorate the agreement.

To calculate the pion $D$ form factor we need first to join smoothly the
function $D^{(\pi a)}(t)$ defined in the vicinity of $t=0$ by
(\ref{GtoA}), (\ref{GpiMIA}) and the suggestion (\ref{G60anz})
with the solution for finite
values of $t$ given by (\ref{GtoA}), (\ref{GpiMIA}), (\ref{G60}).
The smooth join is possible because the function $D^{(\pi a)}(t)$
in the vicinity of zero is defined up to order of $\sim t$. Let us give
some details of the procedure. First, we add to $D^{(\pi a)}(t)$ a cubic
polynom, which vanish at $t\to 0$, with the coefficients $a,b,c$ that are
to be defined by the joint conditions:
\begin{equation}
D^{(\pi\,a)}(t) + a(-t) + b(-t)^2 + c(-t)^3\;,
\label{Dabc}
\end{equation}

We require the form factor (\ref{Dabc}) to be joint smoothly with the form
factor for finite $t$ at a point $t=t_c$.
The coefficients $a,b,c$ and the point $t_c$ can be calculated
unambiguously if the following conditions are satisfied.\\
1. The derivative of the function (\ref{Dabc}) satisfies the following
constraints obtained in \cite{KuS18} (see also \cite{PoS18}):
\begin{equation}
\frac{D^{(\pi\,a)}\,'(0)}{D^{(\pi\,a)}(0)}  =
2.88\,\sim\,3.31\,\mbox{GeV}^{-2}\;.
\label{DsD}
\end{equation}
2. The values of the two functions coincide at the point $t=t_c$, as well
as the values of their first derivatives.\\
3. The form factor (\ref{Dabc})  satisfies the condition
$D^{(\pi\,a)}(t)<0$ that ensures the mechanical stability of the pion.
Note, that the  $D$ form factor defined for finite $t$ does satisfy this
condition. \\
4. For  $t_c$ we choose among all possible points satisfying the
conditions 1--3 the point of maximal absolute value $|t_c|$.
This is necessary for the contribution of singular term $\sim 1/t$ be
minimized at small values of $t$.

\begin{figure}[h!]
\epsfxsize=0.9\textwidth
\centerline{\psfig{figure=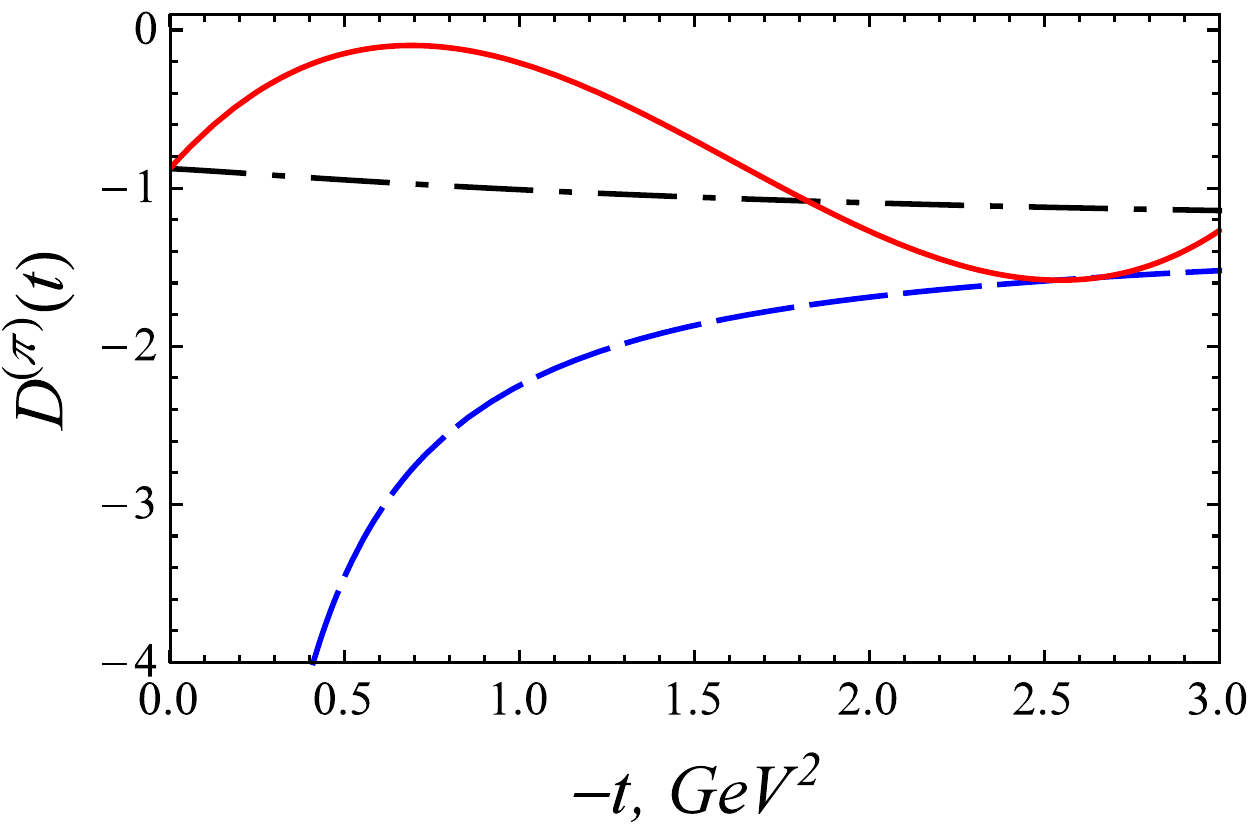,width=9cm}}
\vspace{0.3cm}
\caption{The join of two functions for the pion $D$ form factor for the
middle value from (\ref{DDq}) and the minimal value from (\ref{DsD}).
The full line (red) -- the $D$ form factor (\ref{Dabc}); the dashed
line (blue) - the solution of (\ref{GtoA}), (\ref{G60}), (\ref{GpiMIA});
the dot-dashed line (black) -- $D^{(\pi a)}(t)$; $(-t_c)$ =
2.53\,GeV$^2$.}
\label{fig:3}
\end{figure}

We demonstrate the procedure in fig. \ref{fig:3}, using for the
calculation the minimal value from the interval (\ref{DsD}) and $D_{q} =
-0.1435$.

\begin{figure}[h!]
\epsfxsize=0.9\textwidth
\centerline{\psfig{figure=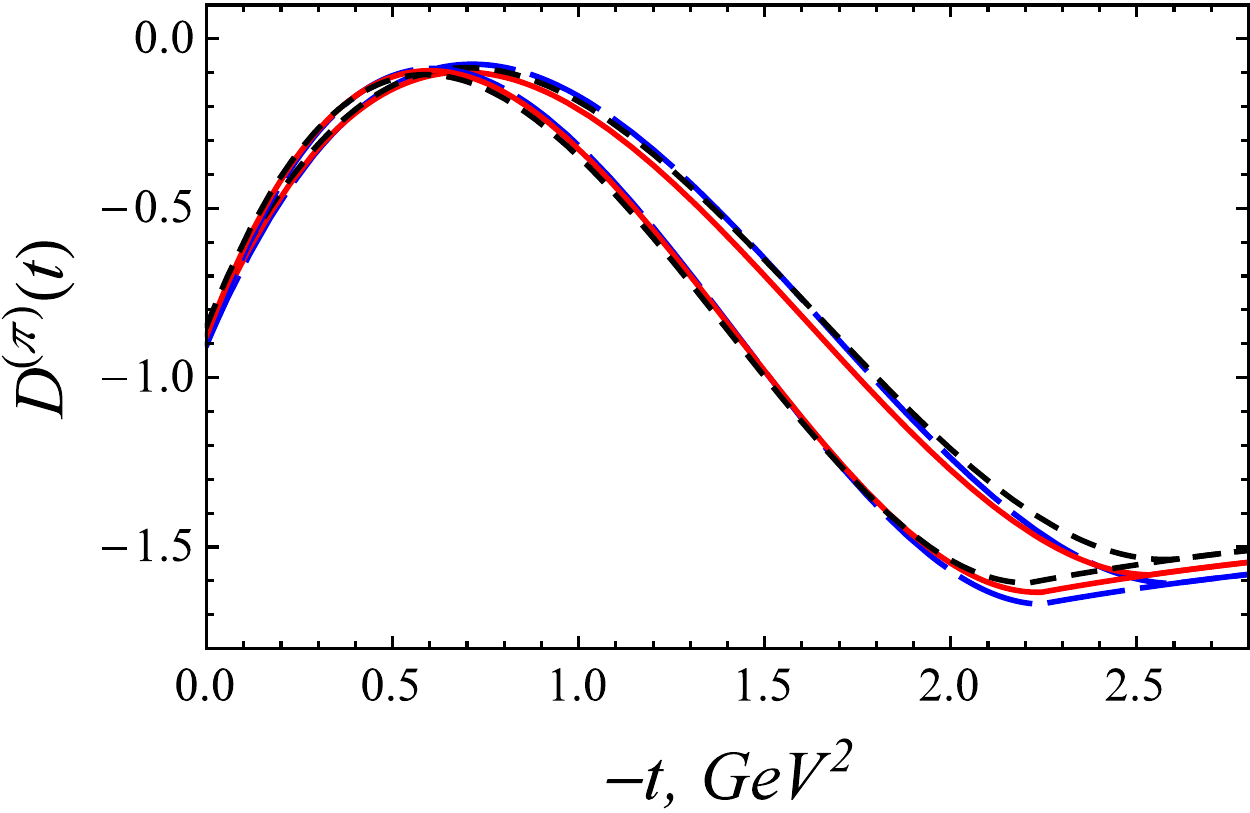,width=9cm}}
\vspace{0.3cm}
\caption{The dependence of the joint $D$ form factor of pion procedure on
the values of the parameter $D_q$ (\ref{DDq}) and on the values from the
interval (\ref{DsD}). The full line (red) -- $D_q=-0.1435$,
long-dashed line (blue) -- $D_q=-0.148$, short-dashed line (margenta) --
$D_q=-0.139$. The upper set of curves at $(-t)\sim$1.5 GeV$^2$ --
for the minimal value from (\ref{DsD}), the lower set -- for the maximal
 value from (\ref{DsD}).}
\label{fig:4}
\end{figure}

We present in fig. \ref{fig:4} the dependence of the procedure on the
values of the parameter $D_q$ (\ref{DDq}) and on the values from the
interval (\ref{DsD}). Fig. \ref{fig:4}  demonstrates the stability of the
procedure.
As can be seen in fig. \ref{fig:4} the result of joining depends weakly on
the value of the quark $D$-term from (\ref{DDq}). However, the parameters
in (\ref{Dabc}) and the point of join $t_c$ do depend on the value from
(\ref{DsD}).

\begin{figure}[h!]
\epsfxsize=0.9\textwidth
\centerline{\psfig{figure=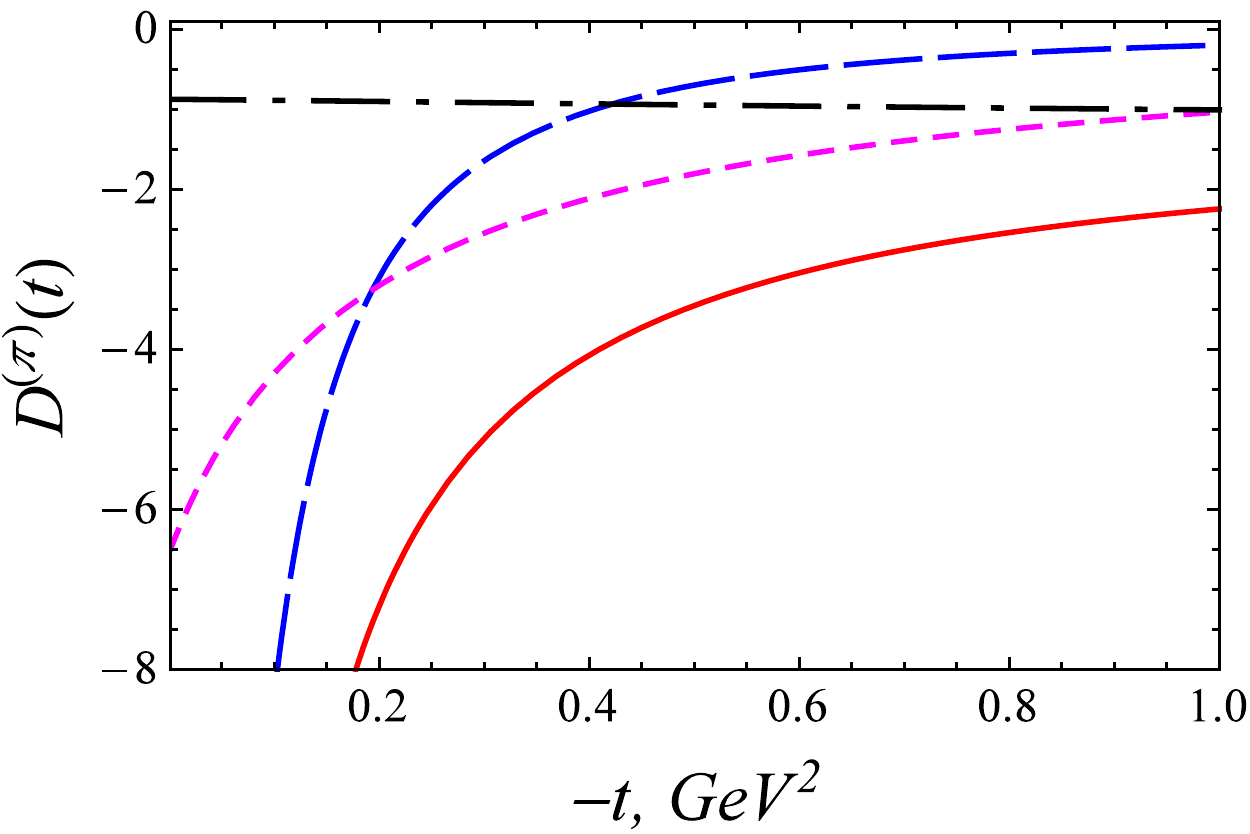,width=9cm}}
\vspace{0.3cm}
\caption{The pion gravitational $D$ form factor calculated for the
parameters (\ref{AqA0}), (\ref{Mb}) and $D_q = -0.1435$ (\ref{DDq}).
The full line (red) -- the total values obtained using  (\ref{GtoA}),
(\ref{G60}), (\ref{GpiMIA}). Long-dashed line (blue) -- the confribution
of quarks $A$ form factors $g^{(q)}_{10}$. Short-dashed line (magenta)--
the spin rotation effect (the contribution of the quark $J$ form factor
$g^{(q)}_{40}$). Dot-dashed line (black) -- the contribution of the quark
$D$ form factor $g^{(q)}_{60}$; this curve coincides with
$D^{(\pi a)}(t)$.}
\label{fig:5}
\end{figure}

The results of calculation of the pion $D$ form factor using the
equations (\ref{GtoA}), (\ref{G60}), (\ref{GpiMIA}), $D^{(\pi a)}(t)$
and the separate contributions of the quark $A$, $J$ and $D$ form factors
are presented in fig. \ref{fig:5}.
One can see from fig. \ref{fig:5} that the singularity in
the pion $D$ form factor at the point $t=0$ is caused by the term
containing the $A$ form factors of
the constituent quarks $g^{(q)}_{10}$ (\ref{g10}), (\ref{G60}). Note also,
that, as well as the pion $A$ form factor, the pion $D$ form factor
contains large contribution of the relativistic spin rotation effect
through the contribution of $J$ form factors of the constituent quarks
$g^{(q)}_{40}$ (\ref{g40}). It is seen that the condition of mechanical
stability of pion $D^{(\pi)}(t)<0$ is fulfilled.

\begin{figure}[h!]
\epsfxsize=0.9\textwidth
\centerline{\psfig{figure=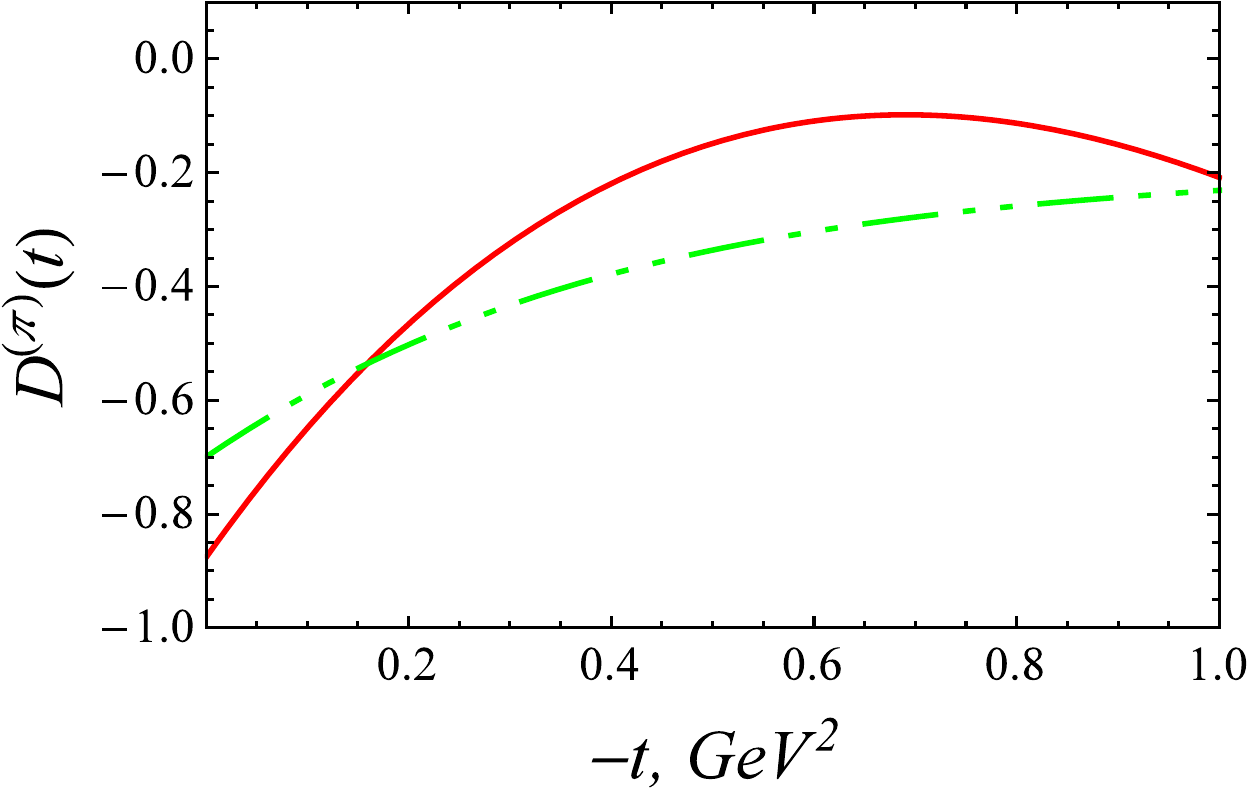,width=9cm}}
\vspace{0.3cm}

\caption{
The pion $D$ form factor calculated  for the quark parameters
(\ref{AqA0}), (\ref{Mb}), $D_q=-0.1435$ (\ref{DDq}) and
the mimimal value of (\ref{DsD}) in comparison with the results of the
paper \cite{KuS18}. The full line (red) -- our result, double-dot-dashed
line (green) -- the pion $D$ form factor from \cite{KuS18}.}
\label{fig:6}
\end{figure}

The pion $D$ form factor calculated with the use of
(\ref{Dabc}) for $D_q=-0.1435$ and the mimimal value of (\ref{DsD}) are
compared with the results of the paper \cite{KuS18} in fig. \ref{fig:6}.

Using the results for the pion GFFs we calculate the mass radius of pion
$$
\langle r^2\rangle_{mass} = 6\,\left.\frac{dA^{(\pi)}}{dt}\right|_{t=0}\;,
$$
and its mechanical radius defihed by (\ref{r2mD}).
To calculate the mass radius we need only the parameters
(\ref{AqA0}), (\ref{Mb}) and so obtain the strictly fixed by our
previous results value $\sqrt{\langle r^2\rangle_{mass}} = 0.1\,$fm.
The chosen interval for $D_{q}$ gives for  the pion mechanical radius
the interval of values $\sqrt{\langle r^2\rangle_{mech}} = 0.82-0.88\,$fm.
It is highly probable that the model with non-point-like quarks will give
larger values for the radii.

Note, that the slopes of the form factors at $t=0$ in fig. \ref{fig:6} are
different. Nevertheless our result for the mechanical MSR as defined above
coincides with that of \cite{KuS18}.

Let us make some remarks concerning a  possibility of comparing  our
results with experimental data.

First, we use an extremely rough approximation - the point-like
constituent quarks. As it was pointed out and argued in detail in
\cite{Kru97}, the accounting for the quark structure, the full quark
form factor, is a necessary part of efficient describing of the
electromagnetic form factors of hadrons. We use here the  simpliest
model aiming to demonstrate that relativistic invariant canonical
parametrization together with MIA in the framework of IF RQM does give
a real possibility of obtaining the pion GFFs. The obtained results
are reasonable and satisfy all standard constraints.

Second, today there are no trustworthy results on pion GFFs unambiguously
extracted from precise experimental data. Although the pion GFFs and
gravitational radii were estimated \cite{KuS18}, the errors of the Belle
measurements are large (even at current stage), and the obtained results
can be affected by the experimental errors. Belle II began data taking
with the much higher luminosity SuperKEKB in 2018, and the precise
measurements of $\gamma^*\gamma\to \pi^0\pi^0$ can be expected since the
 statistic errors are much larger than the systematic errors in the
 previous Belle data \cite{Son19}. One may expect more  quantitative
insights from experiments CLAS at Jefferson Lab \cite{Bis-CLAS}, COMPASS
at CERN \cite{San-COMPASS} and the envisioned future Electron-Ion-Collider
 \cite{Agu19}. \\
\section{Conclusion}
\label{sec: Sec 6}
In this work we extend our relativistic theory of electroweak properties of
composite systems, developed previously, to describe simultaneously the
gravitational structure of hadrons. The approach is based on a version of
the instant-form relativistic quantum mechanics  and makes use of the
modified impulse approximation. We use the general method of the
relativistic invariant parametrizaton of local operators to write the
energy-momentum tensor of particle with an arbitrary spin.
From the point of view of group theory the parameterization
procedure represents the realization of the known Wigner -- Eckart
theorem on the Poincar\'e group. We give general formulae and use for
the actual calculation those for systems of spin $0$(the
pion), spin $1/2$ (the constituent quark) and for the free two-quark system
with total quantum numbers of pion.

To construct the pion GFFs we use
the modified impulse approximation which,
in contrast to the baseline impulse
approximation, is formulated in terms of the form factors
and not in terms of the EMT operator itself.
The pion GFFs
are presented as functionals given by the free two-particle form factors
on the set of the two-quark wave functions of the pion.

We calculate the pion GFFs assuming that the quarks are
point-like. For the two-quark wave function we take the
ground-state wave function of the harmonic oscillator. All but
one parameters of our first-principle model were fixed previously in
works on electromagnetic form factors. The only free parameter, $D_{q}$,
is a characteristic of gravitational form factor of constituent quark,
the quark $D$-term. This parameter is constrained from the pion
mean-square radius despite large uncertainties in the extraction of the
latter from the experimental data through a phenomenological approach.
We calculate the values of the static gravitational characteristics of the
pion and obtain $A$ and $D$ form factors as functions of momentum transfer
up to 1\,GeV$^2$. Note that the new parameter is not used in the
calculation of the $A$ form factor, its value is a direct prediction of
our previous approach.  In the calculation of the $D(t)$ form factor
we use the new parameter and also exploit a special procedure (based on an
{\it ansatz}) to get rid of a singularity at $t=0$.  The important
role of the relativistic effects in the pion gravitational characteristics
is discussed in detail. The calculated gravitational form factors and
gravitational mean-square radii are in a reasonable agreement with the
known results.

The work was supported in part by the Ministry of Science and Higher
Education of the Russian Federation (grant N 0778-2020-0005)(A.K.).

\end{document}